\documentstyle[prd,aps]{revtex}
\begin{document}
\title{Perturbative calculation of the scaled factorial moments in
second-order quark-hadron phase transition within the Ginzburg-Landau
description}
\author{C.B. Yang$^{1,2}$
 and X. Cai$^{1,2}$}
\address{$^1$ Institute of Particle Physics, Hua-Zhong Normal University,
Wuhan 430079, People's Republic of China}
\address{$^2$ Institut f\"ur Theoretische Physik, Freie Universit\"at Berlin,
14195 Berlin, Germany}

\date{\today}
\draft
\maketitle

\begin{abstract}
The scaled factorial moments $F_q$ are studied for a second-order
quark-hadron phase transition within the Ginzburg-Landau description.
The role played by the ground state of the system under low temperature
is emphasized. After a local shift of the order parameter the fluctuations
are around the ground state, and a perturbative calculation for $F_q$
can be carried out. Power scaling between $F_q$'s is shown, and a universal
scaling exponent $\nu\simeq 1.75$ is given for the case with weak
correlations and weak self-interactions.

{{\bf PACS} number(s): 12.38.Mh, 05.70.Fh, 13.85.Hd, 11.30.Qc}
\end{abstract}

\vskip 1cm
\section{Introduction}
It is well-known that ultra-relativistic heavy-ion collision is the unique way
to study the vacuum properties of quantum chromodynamics (QCD)
in the laboratory. In the collisions the kinetic energies of the colliding
particles are converted into thermal ones, and a hot new matter state,
quark-gluon plasma (QGP), might be formed. The system will cool with its
subsequent expanding and will undergo a phase transition from the deconfined
QGP to confined hadrons. Since only the final state particles
in the collisions are observable in experiments, one may be asked to search
for the signals about the phase transition from only those particles.
Since the existence of the phase transition is associated with properties of
the nontrivial chromodynamical vacuum, the study of quark-hadron phase
transition has been a hot point in both particle physics and nuclear
physics for more than a decade. Besides the unique features of QCD
the lack of control of the temperature in the phase
transition distinguishes the problem from the standard critical phenomena
such as ferromagnetism. The nonperturbative nature of hadronization process
in the phase transition precludes at this stage any observable hadronic
predictions from first principles, and some approximate models are used.
One of the models is the Ginzburg-Landau model which can be used as a framework
to calculate various moments of the multiplicity distribution and has been used
in the studies of the scaled factorial moments in both first- \cite{fir-ph}
and second-order \cite{sec-ph} phase transitions, the multiplicity
difference correlators \cite{mdc}, and the multiplicity distributions
in the phase transitions \cite{mul-dis}.

In the Ginzburg-Landau description of a second-order phase transition,
the scaled factorial moments can be expressed as \cite{sec-ph}
\begin{equation}
F_q=f_q/f_1^q, \mbox{\hspace{0.8cm}} f_q={1\over Z}\int {\cal D}\phi
\left(\int_\delta dz\mid\phi\mid^2\right)^q \exp(-F[\phi])
\label{eq:1}
\end{equation}

\noindent with $Z=\int {\cal D}\phi \exp(-F[\phi])$, the free energy functional
$F[\phi]=\int dz [a\mid\phi\mid^2+b\mid\phi\mid^4+c\mid\bigtriangledown
\phi\mid^2]$, $a\propto (T-T_C)$ representing the distance from the critical
point, $b$ and $c$ larger than zero. Here $|\phi|^2$ is associated with
the hadronic multiplicity density of the system, and $\int_\delta dz$ means
integration over a small bin with width $\delta$ in the phase space.
Similar expressions can be derived for other quantities mentioned above.
In all former studies of second-order phase transition the gradient term in
the functional $F[\phi]$ is simply taken to be zero, i.e., the
field $\phi$ is regarded spatially uniform. The spatial integral of the
functional over a two-dimensional bin with size $\delta^2$ is then
$F[\phi]=\delta^2(a\mid\phi^2\mid+b\mid\phi\mid^4)$. This is of course a
very crude approximation. The advantage of such an approximation is
that it turns the functional integration into a normal one. Thus,
the calculation becomes quite easy under the approximation. For $a>0$ the
functional takes its minimum at $|\phi|^2=0$ corresponding to the quark phase,
and for $a<0$ the minimum is at $|\phi|^2>0$ corresponding to the hadron
phase. In all the studies the interested region is for $a<0$. Numerical
results do not show the so-called intermittency behavior, but the
$F$-scaling, $F_q\propto F_2^{\beta_q}$ with universal scaling law $\beta_q
=(q-1)^\nu$, is shown to be valid. The exponent $\nu$ is called as a
universal one in the sense that it is insensitive to the values of the
parameters in the model and that it is completely determined by the structure
of the functional concerned.

The contributions from the gradient term to the moments and to the exponent $\nu$
should be evaluated in some way. Once the gradient term is taken into the
functional, one is faced with serious difficulty in the calculations,
considering the fact that the value of parameter $b$ for the $\phi^4$ term
can be determined in no way from first principles or from experimental output
and may be very large. Even if the parameter $b$ is indeed very small, negative
value of $a$ in our interested region also excludes the possibility of
performing the usual perturbative calculations. The role played by the gradient
term is investigated in \cite{gra-mea} and \cite{gra-sim}. In
\cite{gra-mea} $\phi$ in each bin is still uniform, but the values of $\phi$
in all neighboring bins are taken to be $\phi_0$, field configuration
corresponding to the minimum of ``potential'' $V(\phi)\equiv a\mid\phi\mid^2
+b\mid\phi\mid^4$. So the square of the gradient of $\phi$ is $\delta^{-2}(\phi-
\phi_0)^2$. This approximation also transforms the functional integration
into a normal one. Numerical results show that the universal scaling law
$\beta_q=(q-1)^\nu$ is still valid and that the exponent $\nu$ is almost
the same as without the gradient term. In \cite{gra-sim} the details of
spatial fluctuations of $\phi$ in a bin is simulated by the Ising model for
one-component spins $s$. Each bin is assumed large enough to contain several
spin sites. This time, the exponent $\nu$
depends on the unknown temperature, and, after averaging over the temperature,
$\nu$ is still in the range given in \cite{fir-ph} and \cite{sec-ph}.

Though the simulation in \cite{gra-sim} is convincing, it is for lattice
with one-component spins. In the Ginzburg-Landau model for a second-order phase
transition, the field $\phi$ is a complex number, or in other words,
$\phi$ has two components. At first glimps, the simulation in \cite{gra-sim}
does not correspond to the real problem discussed in the Ginzburg-Landau
model, but as will be explained soon in this paper, it relates to the physics
in an indirect way.

In \cite{gra-pha}, it is tempted to investigate the universality of the
exponent $\nu$, with the spatial fluctuations of the phase angle of the complex
field $\phi$ fully taken into account. As will be shown below, the
contribution from spatial fluctuations of the phase angle of the field
$\phi$ can be evaluated in a complete and rigorous way, and the
integration over the spatial fluctuations of the phase angle of the
field $\phi$ will reduce the problem to one with a one-component field.

The first observation is that all terms except the gradient one in the
functional integral of Eq. (\ref{eq:1}) depend only on $\mid\phi\mid^2$.
Then it is convenient to write the two-component field $\phi$ as a complex
number in the form $\phi= \phi_{\rm R}\exp(i\phi_{\rm I})$. The spatial
fluctuations of the field can
be those of the magnitude $\phi_{\rm R}$ and/or of the phase angle $\phi_{\rm
I}$ (or orientation in an abstract space). The gradient term turns out to be
\begin{equation}
\mid\bigtriangledown\phi\mid^2=(\bigtriangledown \phi_{\rm R})^2
+\phi_{\rm R}^2(\bigtriangledown\phi_{\rm I})^2\ .
\end{equation}

\noindent  Generally, the phase angle $\phi_{\rm I}$ can be in any form,
and the full contribution from its fluctuations
must be evaluated. Fortunately, the integral over $\phi_{\rm I}$ can
be carried out easily since it is of Gaussian form. Then one transforms the
two-fold functional integral into a one-fold one and gets
\begin{equation}
f_q={\int {\cal D}\phi_{\rm R}
\left(\int_\delta dz\phi_{\rm R}^2\right)^q \exp(-F[\phi_{\rm R}])\over
\int {\cal D}\phi_{\rm R}\exp(-F[\phi_{\rm R}])}\ ,
\label{eq:3}
\end{equation}

\noindent with functional $F[\phi_{\rm R}]$ exactly the same form as the
original $F[\phi]$. The important difference between this expression from
Eq. (\ref{eq:1}) is that the functional integral variable in this new
expression is a real function instead of a complex function as in Eq.
(\ref{eq:1}). Then $f_q$ and $F_q$ can be
simulated by a one-component field as in Ref. \cite{gra-sim}.

Now we take the field $\phi_{\rm R}$ (magnitude of $\phi$) uniform, or in
other words, the gradient term of $\phi_{\rm R}$ is omitted. (Calculations
based on this approximation will be referred to mode 2 in this paper.) Based
on the work Ref. \cite{gra-sim} one can drop off the $\bigtriangledown
\phi_{\rm R}$ term, because the problem now is exactly within the scope
of Ref. \cite{gra-sim}, and the conclusions in Ref. \cite{gra-sim}
 encourage us to neglect the spatial fluctuations
of $\phi_{\rm R}$ as long as the universal scaling exponent $\nu$ is concerned.
 Then one gets the factorial moments as functions of variable $x$
\begin{equation}
f_q={\int_0^{\infty} dy y^{2q}\exp(xy^2-y^4)\over \int_0^{\infty} dy
\exp(xy^2-y^4)}\ ,
\end{equation}

\noindent with $x=a\delta^{3/2}/b^{1/4}$. From this expression the scaled
factorial moments $\ln F_q$ can be calculated, and the results are shown as
functions of $-\ln x$ in Fig. \ref{fig1} for $q$ from 2 to 8 within the range
$x\in (0.5, 4.0)$. One can see clearly that no strict intermittency can be
claimed since all $F_q$ approach finite values in the small $x$ limit.
So, no intermittency is shown in the phase transition, as shown in former
studies. More importantly, the power law can be found between $F_q$ and
$F_2$, as shown in Fig. \ref{fig2} with the same data as in Fig. \ref{fig1}.

For the convenience of comparison with former case, we write down the
expressions of the scaled factorial moments without spatial fluctuations
(mode 1 in this paper), which can be read
\begin{equation}
f_q={\int_0^\infty dy y^q
\exp(xy-y^2)\over \int_0^\infty dy \exp(xy-y^2)}\ ,
\end{equation}

\noindent with $x=a\delta/\sqrt{b}$. Numerical results for $\ln F_q$ in this
mode are shown in Fig. \ref{fig3}. In the upper part of the figure $\ln F_q$
are shown as functions of $-\ln x$ for $q$ from 2 to 8 with $x$ in the
same interval $x\in (0.5, 4.0)$, and in the lower part $\ln F_q$ are shown
as functions of $\ln F_2$ with the same data as in upper part. One can see
from upper part of the figure that the general behaviors of $\ln F_q$ as
functions of $-\ln x$ is similar to those in Fig. \ref{fig1},
though the definition of $x$ in this case is different
from that for Fig. \ref{fig1}. The values of $\ln
F_q$ in the two cases are also different. For same value of $x$, $\ln F_q$
in the former case have larger values. This difference is reasonable if
one notices the difference in the definition of variable $x$. What
interests us is the scaling law between $F_q$ and $F_2$. The power law
scaling between $F_q$ and $F_2$ can be seen obviously in the lower part
of Fig. \ref{fig3}, the same as shown in other studies cited in the references.

From Fig. \ref{fig2} and the lower part of Fig. \ref{fig3}, one can
get the scaling exponents
$\beta_q$ for the two different modes by fitting the curves. $\beta_q$ can
also be given analytically. One can expand the expressions for $\ln F_q$ in
the two modes as power series of $x$ in small $x$ limit, and then one gets
the slopes $K_q$ for $\ln F_q$ and $\beta_q=K_q/K_2$. The expressions for
$K_q$ for the two modes in this paper are
\begin{eqnarray*}
&&K_q={\Gamma(q/2+1)\over \Gamma(q/2+1/2)}-q{\Gamma(3/2)\over \Gamma(1)}+(q-1)
{\Gamma(1)\over \Gamma(1/2)}\ \ \ \mbox{for mode 1}\ ,\\
&&K_q={\Gamma(q/2+3/4)\over \Gamma(q/2+1/4)}-q{\Gamma(5/4)\over
\Gamma(3/4)}+(q-1) {\Gamma(3/4)\over \Gamma(1/4)}\ \ \ \mbox{for mode 2}\ .
\end{eqnarray*}

\noindent One can find only a small difference between the exponents
$\nu$ from these two expressions. The results are shown in Fig. \ref{fig4}.
In mode 1
(without spatial fluctuations) $\nu$=1.3335, and in mode 2 (with spatial
fluctuations of the phase angle of the field $\phi$) $\nu$=1.2772. The
exponents obtained from these analytical expressions are very close to
the ones from the fitting. The universal exponents $\nu$ are also very
close to one another and can be regarded
as the same within accuracy $4\%$. Physically, these two modes correspond
to different situations. In mode 1 no spatial fluctuation of $\phi$ is in
the problem, but in mode 2 the spatial fluctuations of the phase angle of
the complex field $\phi$ are fully evaluated. Since these two different
considerations give very close exponents $\nu$, one can
say that the exponent $\nu$ is indeed insensitive to the spatial
fluctuations of the phase angle.

 It is, of course, very
interesting to investigate directly the effect of the term $(\bigtriangledown
\phi_{\rm R})^2$ on the moments, which is the main topic in this paper.

Our second observation is that the final state particles are in a finite
phase space at any high but finite colliding energy. This means that the
fluctuations of the field $\phi$ should not be uniform since the field
must be zero in the region excluded by the conservation laws. Thus there
exists a boundary condition for $\phi_{\rm R}$. For the convenience, we
use $\phi$ instead of $\phi_{\rm R}$ in the following of the paper if no
confusion will be aroused. The boundary condition of $\phi$ is of Dirichlet
type in our problem because of the fact that the particle density out of
a finite region should be zero. In the following, we only discuss a
one-dimensional phase space such as the rapidity, and the boundary
condition can, not losing any generality, be written
as $\phi(0)=\phi(L)=0$, with $L$ the length of the finite phase
space interval. With the gradient term in the functional, the functional
integral can only be calculated perturbatively. But there are two important
differences from the usual perturbations. The first difference is the finite
size of the phase space. The second is the non-positivity of the
coefficient of the Gaussian term in the functional $F[\phi]$. So, some new
techniques are needed which will be discussed in this paper.

The organization of the paper is as follows. In Sec. II we discuss
the ground state of a finite-size system under various boundary conditions.
In Sec. III a new perturbative calculation scheme is proposed with the
effect of local spontaneous symmetry breaking taken into account. In Sec.
IV we calculate the scaled factorial moments perturbatively. Sec. V is for
our main results and conclusions.

\section{Local spontaneous symmetry breaking for finite-size system}

Finite-size effects near critical points have been remained over the past two
decades to be an important topic of the active research both theoretically
and experimentally \cite{fini-bot} in condensed matter physics.
Nowadays, the experimental sample are usually so pure and so well
shielded from perturbing fields that the correlation length can grow up to
several thousand angstroms as the critical point is approached.
When one or more dimensions of a bulk
system is reduced to near or below a certain characteristic length scale,
the associated properties are modified reflecting the lower dimensionality.
It is believed that finite-size effects are precursors of the critical
behavior of the infinite system and can be exploited to extract the limiting
behavior. A central role plays the finite-size scaling behavior predicted by
both the phenomenological \cite{fini-phe} and renormalization group
\cite{fini-ren} theories. Those
theories allowed a systematic discussion of the finite-size effects and,
consequently, form the cornerstone of our current understanding of the way in
which the singularities of an infinite system are modified by the finiteness
of the system in some or all of the dimensions. Of course, the exact form of
scaling functions can't be given in those scaling theories.

In 1985, Br\'{e}zin and Zinn-Justin (BZ) \cite{bz} and Rudnick, Guo and Jasnow
(RGJ) \cite{rgj} developed two field-theoretical perturbation theories for the
calculation of the finite-size scaling functions within the $\phi^4$ model
which corresponds to the Ising model. Most applications of these theories to
three-dimensional systems have been restricted to $T$ higher than the bulk
critical temperature $T_C$ \cite{appl-hi} with a few calculations in region
below $T_C$ \cite{appl-lo}. However, some limitations exist in the theories of
\cite{bz} and \cite{rgj}. As pointed out in the first paper in
\cite{chen-rev}, the theory of BZ is not applicable for $T<T_C$ and the
results from RGJ theory are not quantitatively reliable in the same temperature
region since the coefficients of the Gaussian terms in the integrals are
negative for those temperatures. In \cite{germ-new} the order parameter
is expanded into a sum of eigenfunctions of $\bigtriangledown^2$ for various
boundary conditions. Again, the functional integral is turned out into
a product of normal integrals. But the fluctuations can be evaluated only for
temperature not too far below the critical point.
Authors of \cite{chen-rev} tried to avoid the difficulty
mathematically, but they failed to account for the origin of the difficulty
physically. Although the modified perturbation method in  \cite{chen-rev}
can be used for both $T>T_C$ and $T<T_C$, the calculation is lengthy and can
be done only to first order in practice. Since one does not know the exact
order of values of higher order terms, theoretical results may have large
uncertainty.

It has not been answered that which physical effect causes the failure of direct
perturbative calculations of fluctuations for finite-size systems with
temperature below $T_C$. In our opinion, the real origin of the difficulty
lies in the lack of knowledge about the spontaneously symmetry breaking for
finite-size systems. It is well-known that an infinite system will have
non-zero mean order parameter $\phi_0$, which is called ground state of
the system in this paper since it corresponds to minimum of the Hamiltonian
$H$, if the temperature is below the critical one, and everyone knows that
the difficulty of negative coefficient of the Gaussian term can be overcome
by shifting the order parameter, $\phi\to\phi+\phi_0$. This phenomenon is
known as the spontaneous symmetry breaking because of the fact that $\phi_0$
does not have the same symmetry as $H$ does. This kind of spontaneous
symmetry breaking for infinite system can be called global since the shift
$\phi_0$ is the same constant for every point in the space. For a finite-size
system, such a simple shift of the order parameter does not work
because of the existence of specific boundary conditions for the system.
Anyway, fluctuations of the system, in their own sense, should be around
certain ground state which corresponds to the minimum of the Hamiltonian
$H$, and they can be approximated by Gaussian terms in most cases if
they are not very large. Thus one sees that the ground state plays an
determinative role in the study of fluctuations in the phase transitions
at low temperature. For infinite system, the ground state $\phi_0$ is
constant and can easily be calculated. But for a finite-size system,
the ground state is usually not a constant but depends on the boundary
conditions imposed on it. This is understandable. For infinite system
the ground state is determined completely by the self-interactions of the
field. In other words, the ground state is dictated only by the ``potential'',
and there is no boundary effect. For a finite-size system, however, the
effect of the boundary must be taken into account. For the case with local
interactions, the effect is realized through the gradient term. Thus the
ground state for system with finite size is determined by the gradient
term and the ``potential''. Then the shift of the field at a point depends on
the position in the space. So, the spontaneous symmetry breaking for
finite-size system can be called a local one. Therefore, the solution for
the ground state is non-trivial but necessary, and one has reason to hope
that the difficulty mentioned above for finite-size systems can be overcome
once the ground state is known.

It should be pointed out that all perturbation theories mentioned above are
based on Fourier decomposition of the order parameter. This method is
natural because the decomposition enables one to transform the functional
integral into an infinite product of tractable normal integrals. Although
such a decomposition has simple physical explanation which is very fruitful
for the understanding of properties of infinite systems and can deduce
reliable physical results, as in the case of usual field theories in particle
physics, it brings about a great deal of calculations for finite-size systems.
This is not surprising. As is well-known, quantities complicated
in coordinate space may have simple momentum spectra thus look simple in
momentum space, but those obviously nonzero only in a finite range must
have puzzling momentum spectra. Therefore, for the study of properties of
finite-size systems, calculations in coordinate space might be simpler and
more effective. The point here is that one must calculate the complicated
functional integral which is very difficult to be evaluated directly.

In this section, we first calculate the ground states for a $\phi^4$ model
of a second-order phase transition with one-component order parameter under
various boundary conditions. All the boundary conditions are useful in the
study of condensed matter physics. Then, with the ground states, the
Hamiltonian of the system is reexpressed as Gaussian terms and higher order
perturbations of a locally shifted order parameter. And it is shown that the
perturbative calculation can be done with the new Hamiltonian
for temperatures far below the bulk critical point.

In the $\phi^4$ model for a second-order phase transition in condensed matter
physics with a one-component order
parameter, the partition function can be expressed as a functional
integral of exponential of the Hamiltonian $H$ of the system
\begin{equation}
Z=\int {\cal D}\phi \exp(-H)=\int {\cal D}\phi \exp\left\{-\int d^3\,
{\bf r}\left[{\gamma\over 2}\phi^2+\frac{1}{2}(\bigtriangledown \phi)^2
+\frac{u}{4!}\phi^4 \right]\right\}\ ,
\end{equation}

\noindent in which $\gamma=a^\prime(T-T_C)$, $a^\prime$ and $u$ are
temperature dependent
positive constants, $\phi$ is the order-parameter of the system. In the
following, we are limited only to a film system with thickness $L$.
Since we are interested only in the temperature region $T<T_C$ or $\gamma<0$,
the Hamiltonian $H$ can be standardized by introducing correlation length
$\xi=\sqrt{-1/\gamma}$, new order-parameter $\Psi=\phi/\phi_0$, with
$\phi_0=\sqrt{-6\gamma/u}$ the vacuum expectation of the order parameter
for bulk system, scaled
coordinates ${\bf r}^\prime={\bf r}/L$, and reduced thickness $l=L/\xi$, into
\begin{equation}
H=\int d^3 {\bf r}^\prime {L^3\phi_0^2\over \xi^2}\left[
{1\over 2l^2}(\bigtriangledown^\prime \Psi)^2-\frac{1}{2}\Psi^2+\frac{1}{4}
\Psi^4\right]\ .
\label{eq:fini-2}
\end{equation}

\noindent From this expression one can get the equation for the
ground state by ${\delta H\over \delta\Psi}=0$. The ground state
$\Psi_0(z)$ satisfies
\begin{equation}
{1\over l^2}{d^2\Psi_0\over dz^2}=-\Psi_0+\Psi_0^3\ .
\label{eq:fini-3}
\end{equation}

\noindent In the equation we have used $z$ instead of $z^\prime$
in the range (0, 1) to denote the coordinate along the thickness
direction. Derivatives in other directions do not appear in the
equation since any state with non-zero derivatives in other
directions does not correspond to the minimum of $H$. But if the
system is fully limited in all directions, last equation should
have $\bigtriangledown^2$ in place of $d^2/dz^2$. In
\cite{fini-yang} last equation is solved analytically for
Dirichlet boundary conditions $\Psi(0)=\Psi(1)=0$. The exact
solution is
\begin{equation}
\Psi_0(z)={\sqrt{2}k\over \sqrt{1+k^2}}{\rm sn}(2zF(k), k) \ ,
\label{eq:gs}
\end{equation}

\noindent in which $k$ is determined by $l$ through
$l=2\sqrt{1+k^2} F(k)$. Here, $F(k)$ is the first kind complete
elliptic integral, ${\rm sn}(z,k)$ is elliptic sine function.
Unfortunately, no simple compact solution is found yet for other
boundary conditions. One can easily see that the main obstacle
comes from the nonlinear term $\Psi_0^3$ in the second-order
differential equation of $\Psi_0$ in Eq. (\ref{eq:fini-3}). To
find approximate solutions of $\Psi_0$ for other boundary
conditions, the following method can be used. First of all, we
replace $\Psi_0^3$ in Eq. (\ref{eq:fini-3}) by $\lambda\Psi_0$ and
get a solution satisfying the same boundary condition. For
Dirichlet boundary conditions, the solution is
\begin{equation}
\Psi_0=A\sin\pi z\ , \hbox{\hspace*{0.8cm}with\hspace*{0.5cm}}
\lambda=1.0-\pi^2/l^2.
\end{equation}

\noindent The constant $A$ can be determined by requiring the mean
square of the deviation caused by the replacement, i.e., the
integral $\int_0^1 dz (\Psi_0^3-\lambda\Psi_0)^2$, to be minimum.
Thus one gets
\begin{equation}
\Psi_0(z)=\sqrt{{4\over 3}\left(1-{\pi^2\over l^2}\right)}\sin\pi
z\ . \label{eq:fini-6}
\end{equation}

\noindent Now one can see that the requirement of a minimum
deviation caused by the replacement is equivalent to retaining
$\sin\pi z$ term but neglecting terms with higher frequency in
$\Psi_0^3$. Thus, this approximation is equivalent to the standard
functional variation method. The virtue of this method is that it
can be used simplier and in a step-by-step way. As discussed in
\cite{fini-yang} the ground state is $\Psi_0=0$ if the reduced
thickness $l$ of the film is less than $\pi$. The existence of
minimum reduced thickness of the film implies a shift of the
critical temperature for the finite system from the bulk one. The
exact solutions and the approximate ones are compared in Fig.
\ref{fig5} for $l/\pi$=1.05, 1.10, 1.15, and 1.20. A very good
approximation can be seen. For larger $l$, the same approximative
method can be used further after shift $\Psi_0=
\Psi_0^\prime+\sqrt{4(1-\pi^2/l^2)/3}\sin \pi z$ in Eq.
(\ref{eq:fini-3}). For Neumann boundary conditions,
$\Psi_0^\prime(0)=\Psi_0^\prime(1)=0$, the ground state can also
be obtained in a similar way. The result is
\begin{equation}
\Psi_0(z)=1.0 \hbox{\hspace*{0.5cm} for\hspace*{0.5cm}} T<T_c
\end{equation}

Then one can consider mixed boundary conditions $\Psi_0(0)=0,
\Psi_0^\prime(1)=0$.
The first order approximation of the solution for ground state is
\begin{equation}
\Psi_0(z)=\sqrt{{4\over 3}\left(1-{\pi^2\over
4l^2}\right)}\sin{\pi z\over 2}\ \mbox{\hspace*{0.5cm}} \hbox{for\
\ \ } l\ge \pi/2\ .
\end{equation}

As a final example, we give the ground state for periodic boundary
condition $\Psi_0(z)=\Psi_0(1+z)$. The ground state is
\begin{equation}
\Psi_0(z)=1.0 \hbox{\hspace*{0.5cm} for\hspace*{0.5cm}} T<T_c
\end{equation}

Though the ground state for periodic and Neumann boundary conditions
are the same the fluctuations of the fields in the two case are different.
It should be pointed out that $-\Psi_0$ is also a ground state of the
system. Then the fluctuations of the system can be around either $\Psi_0$ or
$-\Psi_0$. This is the copy for finite-size systems of spontaneous symmetry
breaking in $\phi^4$ model. The significant difference from the usual
spontaneous symmetry breaking is that the ground state is usually not a
constant and depends on the boundary conditions.  So that we have a local
spontaneous symmetry breaking in this paper. With the ground state $\Psi_0$,
one can locally shift the order parameter $\Psi= \Psi^\prime+\Psi_0$,
then the Hamiltonian $H$ turns out to be
\begin{equation}
H=H[\Psi_0]+{L^3\phi_0^2\over \xi^2}\int d^3 {\bf r}{1\over
2}\left[{1\over l^2}(\bigtriangledown
\Psi^\prime)^2-{\Psi^\prime}^2+ 3\Psi_0^2{\Psi^\prime}^2
+2\Psi_0{\Psi^\prime}^3+{1\over 2}{\Psi^\prime}^4\right] \ .
\label{eq:basic}
\end{equation}

\noindent In this expression, $H[\Psi_0]$ has the same form as $H[\Psi]$ in
Eq. (\ref{eq:fini-2}) with $\Psi_0$ in place of $\Psi$. Now the quadratic part
of fluctuation $\Psi^\prime$ is positive definite for $l$ larger than a
characteristic length, or for temperature enough below the critical
point. Then one sees that the new Hamiltonian can be safely used to
calculate perturbatively fluctuations at low temperature region for
finite-size systems. Then the difficulty of the negative coefficients of the
Gaussian terms is avoided after the effects of local spontaneous
symmetry breaking are taken into consideration.

\section{Perturbative theory for finite-size system under $T\ll T_C$}

From Eq. (\ref{eq:basic}), a new perturbative theory can be
developed for finite-size system with local spontaneous symmetry
breaking. First of all, one can introduce for an one dimensional
system a generating functional $Z[J]$
\begin{equation}
Z[J]=\int {\cal D}\phi \exp(-H+\int dz J\phi)\ .
\end{equation}

\noindent The generalization to more general cases is obvious. Up
to an unimportant constant factor, the generating functional for a
one-dimensional system can, in a standard way, be written as
\begin{equation}
Z[J]=\exp(\lambda_1 \int dz J\Psi_0) \exp\left\{-\lambda\int
dz\left[ \Psi_0\left({\delta\over \lambda_1\delta
J}\right)^3+{1\over 4} \left({\delta\over \lambda_1\delta
J}\right)^4\right]\right\} \exp
\left[\frac{1}{2}\frac{\lambda_1^2}{\lambda} \int dz dy
J(z)G(z,y)J(y) \right]\ , \label{eq:per1}
\end{equation}

\noindent with $\lambda_1=L\sqrt{6|\gamma|/u}=L\phi_0$,
$\lambda=6L\gamma^2/u$. In last equation, the Green's function
$G(z,y)$ satisfies
\begin{equation}
\left[-{1\over l^2}{d^2\over dz^2}-1+3\Psi_0^2(z)\right]
G(z,y)=\delta(z-y)\ . \label{eq:per2}
\end{equation}

\noindent The first factor in the generating functional shows a
great difference between present theory and the usual ones that
there exists a nontrivial solution for the classical equation
$\delta H/\delta \phi=J$ for $J=0$. For systems with higher
dimension $d>1$ the only changes are with $L^d$ in place of $L$ in
the expressions for parameters $\lambda$ and $\lambda_1$ and with
$\bigtriangledown^2$ in place of $d^2/dz^2$ in last equation. The
Green's function $G(z,y)$ describes fluctuations in the full space
and determines how the fluctuations at different points are
correlated. If one can get the solution for $G(z,y)$ for higher
dimensional system, the fluctuations can be evaluated in the same
way as for one dimensional system. Thus in the following we do not
distinguish one and higher dimensional systems, and $dz$ is used
to represent the integral element over a volume in certain space.
From Eq. (\ref{eq:per1}), it can be seen that each Green's
function $G$ is associated with a factor $1/\lambda$. $\lambda_1$
can be regarded as a factor associated with the external source
field $J$. Since the derivative terms in the second factor in Eq.
(\ref{eq:per1}) with respect to the external source field $J$ will
generate terms with more factors of $G$ in the generating
functional, the contribution of them  is small if the parameter
$\lambda$ is big enough. Then those terms in the generating
functional can be regarded as perturbations. From the expression
of $\lambda$ it is clear that a large $\lambda$ is equivalent to a
small $u$ for fixed $L$ and $\gamma$. Thus the condition of a
large $\lambda$ is consistent with that in usual perturbation
theory. Then one has all the four ingredients diagrammatically
represented in Fig. \ref{fig6} for the perturbative calculations
with the Feynman rules:
\begin{description}
\item\hspace*{1cm} (a) the ground state: $\lambda_1\Psi_0(z)$,
\item\hspace*{1cm} (b) the Green's function (propagator):
${\lambda_1^2\over \lambda} G(z, y)$,
\item\hspace*{1cm} (c) three-line vertex: $-{\lambda\over
\lambda_1^3}\int dz \Psi_0(z)$,
\item\hspace*{1cm} (d) four-line vertex: $-{\lambda\over 4\lambda_1^4}\int dz$.
\end{description}

\noindent Using these ingredients all physical quantities can be calculated.
For example, to the first order of the perturbations, one has
\begin{eqnarray*}
&&\langle \Psi(z)\rangle = \Psi_0(z)-{3\over \lambda}\int du
\Psi_0(u) G(u,u)G(u,z)\ ,\\ &&\langle
\left(\Psi(z)-\Psi_0(z)\right)\left(\Psi(y)-\Psi_0(y)\right)\rangle
= {1\over \lambda}G(z,y)-{3\over \lambda^2} \int du
G(z,u)G(u,u)G(u,y)\ .
\end{eqnarray*}

\noindent Here, the symbol $\langle\cdots\rangle$ represents the
average over the fluctuations, the range of the integral over $u$
is (0, 1).

A most important feature of the perturbation theory is that all the
calculations can be done in coordinate space. Once the non-trivial
ground state $\Psi_0$ is known, one can get the Green's function (propagator)
$G(x,y)$ from Eq. (\ref{eq:per2}), and other quantities can be obtained
from Eq. (\ref{eq:per1}) by directly taking derivatives with respect to the
external source field $J$. This scheme can be used in calculating properties of
finite-size systems in condensed matter physics for temperatures $T\ll T_c$.

Next section will calculate the scaled factorial moments in a second-order
quark-hadron phase transition as an example
of the applications of the perturbation theory.

\section{The scaled factorial moments in the Ginzburg-Landau model}

Now we turn to the calculation of the scaled factorial moments
$F_q$ in Eq. (\ref{eq:1}) in a second-order quark-hadron phase
transition within the Ginzburg-Landau description. In this
description, the free energy functional $F[\phi]$ is in place of
the Hamiltonian $H$ in last two sections. After integrating over
the phase angle of the field the functional remains the same form
with a real $\phi_{\rm R}$ in place of the complex $\phi$ as
discussed before. Although there are very important differences
between normal phase transitions in condensed matter physics and a
quark-hadron one, the mathematical form in the Ginzburg-Landau
description for them is the same. In the Ginzburg-Landau
description for a quark-hadron phase transition, the integral
variable $z$ is not in coordinate space but represents a
collection of measurable quantities such as rapidity and azimuthal
angle etc. In the following, $z$ is identified to the rapidity.
For such an one dimensional system, the local spontaneous symmetry
breaking is also given as in Sec. II. A generating functional can
also be introduced in the same way as in last section. The only
changes are the expressions for the parameter $\lambda$,
$\lambda_1$, and $l$. Here we only mention the expression for $l$.
In present case, the correlation length is $\xi=\sqrt{c/|a|}$, so
$l=L\sqrt{|a|/c}$. The parameter $c$ has a simple physical
meaning. From the free energy functional one sees that the
correlation between fields at different points is realized by
means of the gradient term. If $c$ is small there is weak
correlation between the fields at different points. Thus the
effective length $l$ can be used to measure the strength of the
correlations for fixed $L$ and $|a|$. For a system at fixed
temperature $c$ is small if there is weak correlation, and vice
versa. When $l\to\infty$, one may expect that the influence of
correlation can be neglected and that the effect of boundary
condition can be neglected. In the calculation of the scaled
factorial moments, the factor $\lambda_1$ will be cancelled. So
$\lambda_1$ can be taken to be 1.0 in present calculations. For
any parameter $l$ the scaled factorial moments can be rewritten
from Eq. (1) as
\begin{equation}
F_q=f_q/f_1^q\ ,\ \ \ f_q=\prod_{i=1}^q \int_\delta dz_i
{\delta^2\over \delta J^2(z_i)} {Z[J] \over Z[0]}\ .
\label{eq:fact1}
\end{equation}

\noindent In this expression, $\int_\delta dz$ represents an
integral over a range of length $\delta$. In our calculation, the
integral range is chosen around the center of the interval (0, 1),
or in other words, in the range (1/2-$\delta$/2,1/2+$\delta$/2).
As discussed in the second last paragraph in Sec. I, the boundary
condition for our case is of Dirichlet type. So the ground state
$\Psi_0$ is given by Eq. (\ref{eq:gs}) and $G(z, y)$ is calculated
from Eq. (\ref{eq:per2}).

\subsection{Zero order approximation for $f_q$}

We first calculate the scaled factorial moments $F_q$ in a second-order
quark-hadron phase transition at the zero order (or tree-level)
approximation to the functional Eq.
(\ref{eq:per1}). At this level the second factor in Eq. (\ref{eq:per1})
gives a factor 1.0. In the expressions of $f_q$ there are contributions
from $q$-particle correlations represented diagrammatically by connected
diagrams in Fig. \ref{fig7} and the contributions from fewer particle
correlations which can be represented by products of disconnected diagrams.
We denote $f_q^c$ the contributions to $f_q$ from connected diagrams which
give the contribution from the pure $q$-particle correlations to $f_q$. Then
the factorial moments $f_q$ at tree-level can be written as
\begin{equation}
\begin{array}{c}
f_1^{\rm tree}=f_1^c\ ,\cr f_2^{\rm tree}=(f_1^c)^2+f_2^c\ ,\cr
f_3^{\rm tree}=(f_1^c)^3+3f_1^cf_2^c+f_3^c\ ,\cr f_4^{\rm tree}=
(f_1^c)^4+6(f_1^c)^2f_2^c +4f_1^cf_3^c+3(f_2^c)^2+f_4^c\ ,\cr
f_5^{\rm
tree}=(f_1^c)^5+10(f_1^c)^3f_2^c+10(f_1^c)^2f_3^c+5f_1^cf_4^c
+10f_2^cf_3^c +f_5^c\ , \cr  f_6^{\rm
tree}=(f_1^c)^6+15(f_1^c)^4f_2^c+20(f_1^c)^3f_3^c
+15(f_1^c)^2f_4^c+6f_1^cf_5^c+10(f_3^c)^2+15(f_2^c)^3+15f_2^cf_4^c+
60f_1^cf_2^cf_3^c+f_6^c\ ,\cr \cdots\ \ .
\end{array}
\end{equation}

\noindent For the connected contributions to $f_q^{\rm tree}$, there are only
two topologically different diagrams, as shown in Fig. \ref{fig7}. For the first
type diagram with two crosses representing the ground state, the number of
identical terms is $N_q^1=2^{q-1}q!$. The factor $q!$ comes from the exchange
symmetry of the $q$ particles, $2^q$ from the two-lines from each point
representing a particle, and a factor 1/2 from the identities of terms
with reversal order of the $q$-points. For the second type of diagrams with
no cross, the number is $N_q^2=N_q^1/q=2^{(q-1)}(q-1)!$. To
calculate the diagrams, it would be useful to define
\begin{equation}
g_i(z,y)=\int_\delta dx_1\ dx_2\cdots\ dx_i\ G(z,x_1)
G(x_1,x_2)\cdots G(x_i,y)\ ,
\end{equation}

\noindent which satisfies a recursive relation
\begin{equation}
g_i(z,y)=\int_\delta du g_{i-1}(z,u)G(u,y)=\int_\delta
 dz G(z,u)g_{i-1}(u,y) \ .
\end{equation}

\noindent Then the contribution from each connected diagram for $f_q$
can be written as
\begin{eqnarray*}
&\hbox{first \ \ disgram:}\hspace*{0.5cm} & \left({1\over
\lambda}\right)^{q-1} \int_\delta dz\ dy
\Psi_0(z)g_{q-2}(z,y)\Psi_0(y)\ ,\\ &\hbox{second
disgram:}\hspace*{0.5cm} & \left({1\over \lambda}\right)^q
\int_\delta dz g_{q-1}(z,z)\ .
\end{eqnarray*}

\noindent So that
\begin{equation}
f_q^c={2^qq!\over \lambda^q}\left[ {\lambda\over 2}\int_\delta dz
dy\ \Psi_0(z) g_{q-2}(z,y)\Psi_0(y)+\frac{1}{2} \int_\delta dz
g_{q-1}(z,z) \right]\ . \label{eq:tree1}
\end{equation}

\subsection{First order approximation for $f_q$}

Now we discuss $f_q$ at the first order (1-loop level) approximation
of the second-factor in the functional of Eq. (\ref{eq:per1}). At this
approximation, the factor from the second term of the equation is
\begin{eqnarray*}
Z_1[J]=1-{1\over \lambda}\int dz \left\{ \Psi_0(z)\left[3G(z,z)(GJ)_z+{1\over
\lambda}(GJ)_z^3\right]+\frac{1}{4}\left[ 3G^2(z,z)+{6\over\lambda}G(z,z)(GJ)_z^2
+{(GJ)_z^4\over\lambda^2}\right]\right\}\ ,
\end{eqnarray*}

\noindent in which $(GJ)_z\equiv \int du G(z,u)J(u)$\ . From the
functional at this approximation $$Z[J]=Z_1[J]\exp(\int dz
J\Psi_0) \exp \left[\frac{1}{2\lambda}\int dz dy
J(z)G(z,y)J(y)\right]$$ the factorial moments $f_q$ can be
directly calculated by using Eq. (\ref{eq:fact1}). There are many
terms contributing to $f_q$, among which the most interesting
terms are those represented by connected diagrams in Fig.
\ref{fig8} with one bulb which is the vertex for the perturbative
interactions. The sum of the contributions from the diagrams to
$f_q$ will be denoted by $f_q^{\rm loop}$ in this paper. Then up
to the first order approximation of the generating functional, the
factorial moments $f_q$ are
\begin{equation}
\begin{array}{l}
f_1=f_1^{\rm tree}+f_1^{\rm loop}\ ,\\ f_2=f_2^{\rm
tree}+2f_1^{\rm tree}f_1^{\rm loop}+f_2^{\rm loop}\ ,\\
f_3=f_3^{\rm tree}+3f_2^{\rm tree}f_1^{\rm loop}+3f_1^{\rm
tree}f_2^{\rm loop} +f_3^{\rm loop}\ ,\\ f_4=f_4^{\rm
tree}+4f_3^{\rm tree}f_1^{\rm loop}+6f_2^{\rm tree}f_2^{\rm loop}+
f_1^{\rm tree}f_3^{\rm loop}+f_4^{\rm loop}\ ,\\ f_5=f_5^{\rm
tree}+5f_4^{\rm tree}f_1^{\rm loop}+10f_3^{\rm tree}f_2^{\rm loop}
+10f_2^{\rm tree}f_3^{\rm loop}+5f_1^{\rm tree}f_4^{\rm
loop}+f_5^{\rm loop}\ ,\\ f_6=f_6^{\rm tree}+6f_5^{\rm
tree}f_1^{\rm loop}+15f_4^{\rm tree} f_2^{\rm loop}+20f_3^{\rm
tree}f_3^{\rm loop}+15f_2^{\rm tree}f_4^{\rm loop} +6f_1^{\rm
tree}f_5^{\rm loop} +f_6^{\rm loop}\ ,\\ \cdots\ \ \ .
\end{array}
\end{equation}

For the perturbative calculation to have high accuracy, we choose
the parameter $\lambda$ large enough to guarantee the following
two conditions: (1) $\int dz |\langle\Psi(z)\rangle-\Psi_0(z)|$,
the integral of absolute deviation of the mean value of the order
parameter $\Psi$ from $\Psi_0$ is not larger than 0.05; (2)
$|Z[0]-1|$ is no more than 0.05. These two conditions ensure the
contributions from higher order terms from the second-factor in
Eq. (\ref{eq:per1}) can be safely neglected. So our calculations
are limited to only the first order approximation. Of course,
higher order approximation can be made without difficulty in
principle, only with more diagrams drawn and evaluated. In
numerical calculation, $\delta$ is chosen for $-\ln\delta$ in the
range (1,4).

\section{Main results and discussions}

As discussed in last section, we choose the parameter $\lambda$ to be a large
number to ensure the small influence of the perturbations. A large
$\lambda$ corresponds to a small correlation function $G(x,y)/\lambda$. So
in the Ginzburg-Landau model for a second-order phase transition under some
choice of the parameters, there is weak correlation between the fields at
different points together with weak self-interactions.
Due to the choice of a large $\lambda$ ($\lambda_1=1.0$) the ground
state $\Psi_0$ (whose square is the hadronic density at the state) will play a
dominant role in Eq. (\ref{eq:per1}) for $l$ large enough, and the Gaussian
and higher order fluctuations can only bring about some small corrections
to the generating functional. Then with the choice of $\lambda$ we are dealing
with a case with small fluctuations. Because of the large value of
$\lambda$ the first term in the brackets of Eq. (\ref{eq:tree1}) plays an
important role if the ground state $\Psi_0$ is obviously nonzero for larger
parameter $l$. Since the powers before the brackets of  Eq. (\ref{eq:tree1})
will be cancelled, the order of the ratios $f_q^c/(f_1^c)^q$ is
$\lambda^{-(q-1)}$, thus very small.  $f_q^{\rm loop}/(f_1^c)^q$ have the
order $\lambda^{-q}$, even smaller. Then the scaled factorial moments $F_q$
are very close to 1.0. This expectation is confirmed in numerical calculations.
Numerical results show that $\ln F_q$, though very small, have quite
complicated behaviors. They increase for $-\ln \delta$ within (1.5, 2.5) and
then decrease with the increase of $-\ln \delta$, as shown in Fig. \ref{fig9}
for parameter $l=2.63\pi$. Thus there is no intermittency in the phase
transition. For other choices of $l/\pi\gg 1$ similar results can be obtained.
A more important and more interesting phenomenon is the power scaling
between $F_q$ and $F_2$, $F_q\propto F_2^{\beta_q}$, which can be expected from
the similar behaviors of $\ln F_q$ in Fig. \ref{fig9} and are shown in Fig.
\ref{fig10} with the same data as in Fig. \ref{fig9}. $\beta_q$ can be obtained
easily from a linear fitting to the curves in Fig. \ref{fig10}. As in former
studies of $F_q$ in Refs. \cite{fir-ph,sec-ph} in the phase transitions, $\beta_q$ satisfies
a universal scaling law
\begin{equation}
\beta_q=(q-1)^\nu\ , \hbox{with\ \ \ }\nu=1.7539\ \ \ \hbox{for}\ \
l/\pi=2.63\ ,
\end{equation}

\noindent
as shown in Fig. \ref{fig11}. In this case the universal exponent $\nu$
depends only on the value of parameter $l$ which is a function of
parameters $|a|$ for the temperature and $c$ for the correlation strength.
The dependence of the exponent $\nu$ on temperature is consistent with Ref.
\cite{gra-sim}. But the exponent $\nu$ is very different from those
exponents given in former studies. The discrepancy is caused from the
different assumptions made in former studies and in present one.
In former studies, the effect of the gradient term is neglected, but the $\phi^4$ term (which describes
the self-interaction) is emphasized. In those studies, the factorial moments
$f_q$ can be written as
\begin{equation}
f_q=\int^\infty_0 dy y^q\exp(xy-y^2)/\int^\infty_0 dy \exp(xy-y^2)\ ,
\end{equation}

\noindent in which $x$ is a parameter representing the bin width. From this
expression one can discover that the $\phi^4$ term, corresponding to the
$-y^2$ term in the exponentials, is very important and cannot
be treated as perturbation for any parameter $x$. It is the term
that makes the integrals finite. In present calculations, the role
played by the $\phi^4$ term is much less important. Its function is
to provide a nontrivial ground state $\Psi_0$ around which the fluctuations
are. Then that term is treated as a small perturbation and is very weak
indeed with our choice of parameter $\lambda$. In former studies the
fluctuations are arouns $\phi_0=0$. Then the discrepancy
between present study and former ones can be understood because they belong
to different physical regimes. Former studies are in the nonperturbative
regime with trivial ground state, but present study in the perturbative one
with a non-trivial ground state.

The dependence of the universal exponent $\nu$ on the parameter $l$ is also
studied for $l/\pi>1$ in which there exists a non-trivial ground state. The
result is shown in Fig. \ref{fig12}. For long correlation length ($l/\pi$
a little larger than 1.0) the fluctuations in neighboring bins are correlated.
For these $l$ the values of $\Psi_0$ are also small, so the two terms in
the bracket in Eq. (\ref{eq:fact1}) may have comparable contributions
to $f_q$. In this region $\nu$ is
quite large (about 2). With the increase of $l$ the correlation between
the fluctuations in neighboring bins becomes weaker and weaker, and
the exponent $\nu$ decreases first rapidly and then slowly. When
$l/\pi>2.5$ $\nu$ approaches a constant, about 1.75. The constant can be
anticipated by considering a case with the very weak correlations among
particles more than 2 (considering the factor $1/\lambda$ accompanied with
the Green's function $G(x,y)$). Then if only the effects of a weak
two-particle correlation is considered, one has
$$f_q=(f_1^c)^q+C_q^2 (f_1^c)^{q-2}f_2^c\ ,$$
and then
$$F_q=1.0+C_q^2 f_2^c/(f_1^c)^2\ .$$
Here $C_m^n$ are the binormial coefficients. Since the ratio $f_2^c/(f_1^c)^2$
is assumed very small one gets $\ln F_q\simeq C_q^2 f_2^c/(f_1^c)^2$, so
the linear relation between $\ln F_q$'s can be verified, and one can get
$$\beta_q=C_q^2=q(q-1)/2\ .$$ From this expression, one gets the exponent
$\nu=1.7550$, very close to the one obtained in this paper.

As a summary, the spatial correlation of the fluctuations in a second-order
quark-hadron phase transition is considered in this paper within the
Ginzburg-Landau description. We deal with a case with finite phase space
and with negative coefficient of the Gaussian term in the functional. Because of
the finite size of the space, calculations in usual space are simpler and more
effective. Due to the negative coefficient of the Gaussian term in the
functional a local spontaneous symmetry breaking (or non-trivial ground state)
exists for finite size system. We emphasize on the importance of the ground
state of the system, which is a version of spontaneous symmetry breaking for
finite-size systems. Then a new perturbation scheme
is developed which is expected to be applicable in the low temperature region
in the $\phi^4$ model for second-order phase transitions in condensed matter
physics. Then as an example of the applications, the scaled factorial
moments $F_q$ in a second-order quark-hadron phase transition are calculated
perturbatively. Power scaling laws between $F_q$'s are shown and a universal
exponent $\nu$ is given.

\acknowledgments
   This work was supported in part by the NSFC, the MSE and the
   Hubei-NSF in China and the DFG in Germany.  Authors thank
   Professor T. Meng for his kind hospitality during their visits in
   FU Berlin, and one of them (C.B. Yang) is grateful for fruitful
   discussions with Professor R. C. Hwa.

\newpage
\begin{center}
{\Large Figure Captions} \end{center}

\begin{figure}
\caption{Dependences of $\ln F_q$ on the bin width $-\ln x$ after
the contribution from spatial fluctuations of the phase angle of
the f ield fully taken into account (mode 2). Curves from lower to
upper are for $q$ from 2 to 8, respectively.} \label{fig1}
\end{figure}

\begin{figure}
\caption{ Scaling behaviors of $\ln F_q$ vs $\ln F_2$ for the same
data as in Fig.1.} \label{fig2}
\end{figure}

\begin{figure}
\caption{ Upper part: $\ln F_q$ as functions of $-\ln x$ without
spatial fluctuations (mode 1); Lower part: Scaling behaviors
between $\ln F_q$ and $\ln F_2$, with the same data as in the
upper part.}\label{fig3}
\end{figure}

\begin{figure}
\caption{Scaling behaviors of $\ln \beta_q$ as function of $\ln
(q-1)$ for the two modes.}\label{fig4}
\end{figure}

\begin{figure}
\caption{Comparison between exact solutions and approximate ones
for Eq. \ref{eq:fini-3} under Dirichlet boundary conditions for
$l/\pi$=1.05, 1.10, 1.15, and 1.20. The solid curves correspond to
exact solutions, dotted curves are drawn according to Eq.
(\ref{eq:fini-6}).} \label{fig5}
\end{figure}

\begin{figure}
\caption{Feynman diagram representations for (a) the ground state,
(b) the propagator, (c) three-line vertex, and (d) four-line
vertex.} \label{fig6}
\end{figure}

\begin{figure}
\caption{Connected zero-order diagrams for the contributionsr
 to $f_q$. In the diagrams the number of dots is equal to $q$ and an
integral over the coordinate in a range with length $\delta$ is implied.
So (a) and (b) are for $f_1$, and (c) and (d) are for $f_q$ with $q$ dots
in the diagrams.}
\label{fig7}
\end{figure}

\begin{figure}
\caption{Connected first-order diagrams for the contributions to
$f_q$ for $q$=1, 2, 3, 4, 5, 6, respectively.} \label{fig8}
\end{figure}

\begin{figure}
\caption{Dependences of the scaled factorial moments $\ln F_q$ on
the bin width $-\ln\delta$ from 1.0 to 4.0 for parameter
$l=2.63\pi$ for $q$=2, 3, 4, 5, and 6.} \label{fig9}
\end{figure}

\begin{figure}
\caption{Power scaling between $F_q$'s with the same data as in
Fig. \ref{fig5}.} \label{fig10}
\end{figure}

\begin{figure}
\caption{Universcal scaling between $\beta_q$ and $(q-1)$ for
parameter $l=2.63\pi$.} \label{fig11}
\end{figure}

\begin{figure}
\caption{Dependence of the universal exponent $\nu$ on parameter
$l/\pi$.} \label{fig12}
\end{figure}
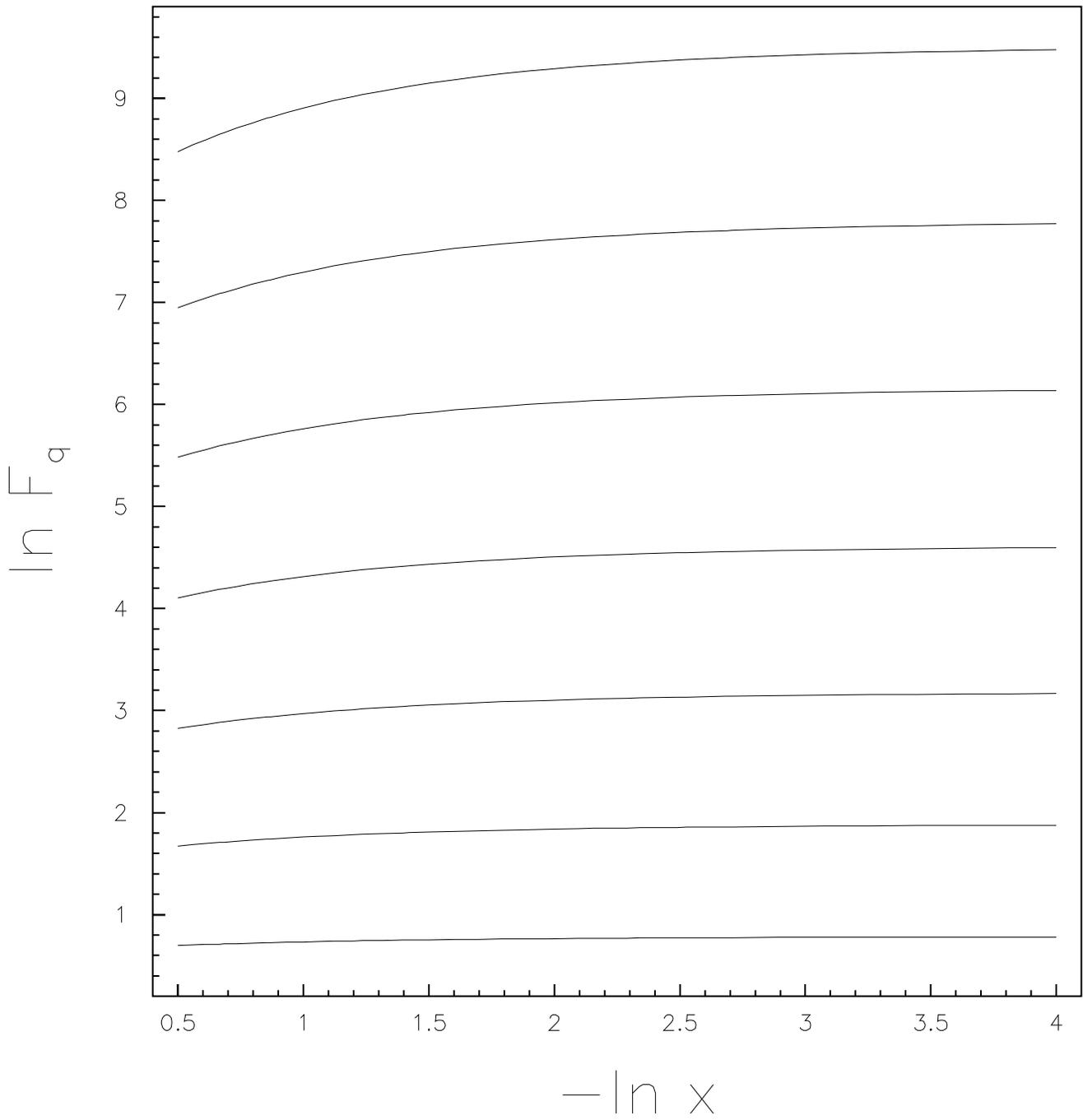
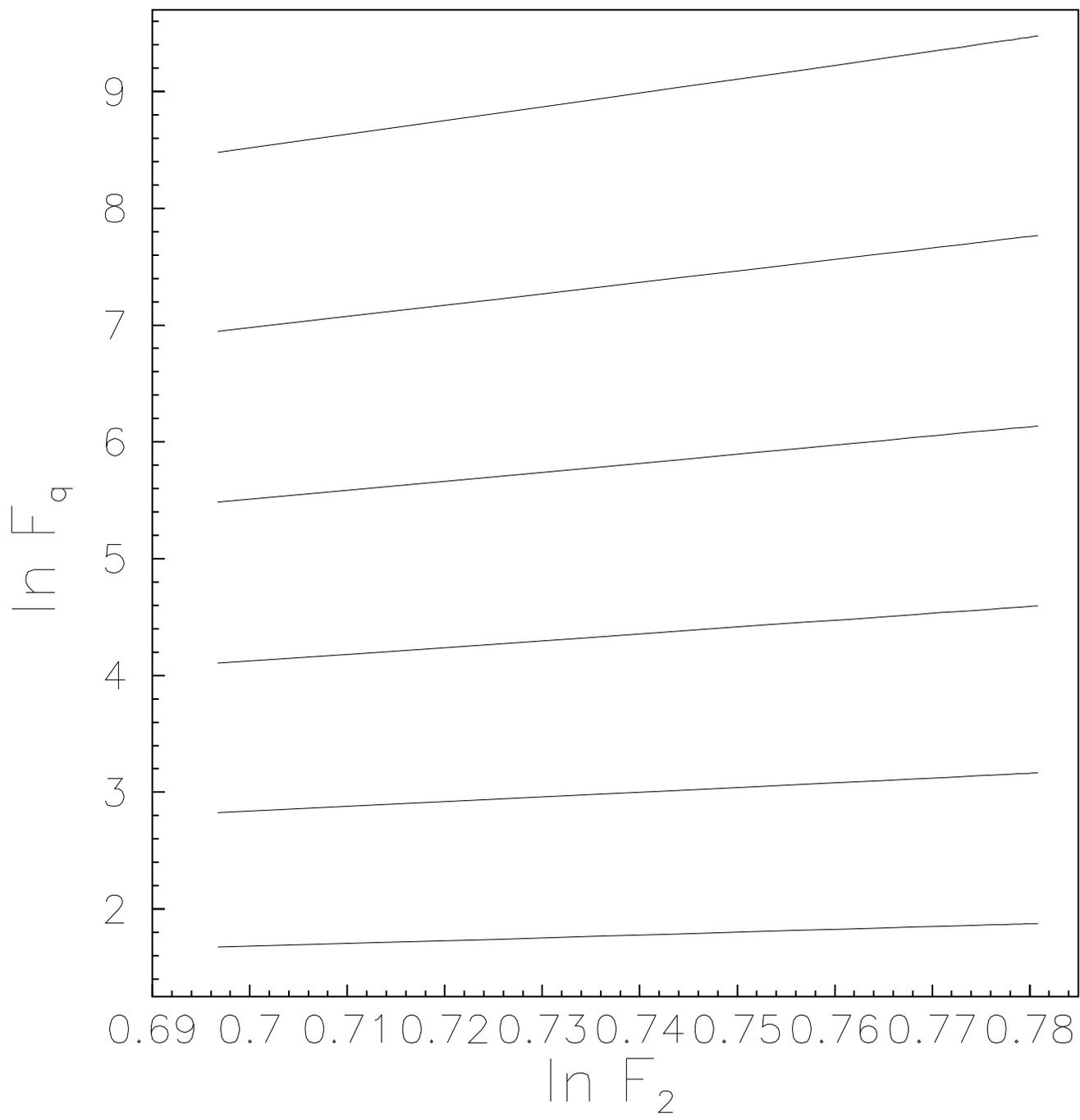
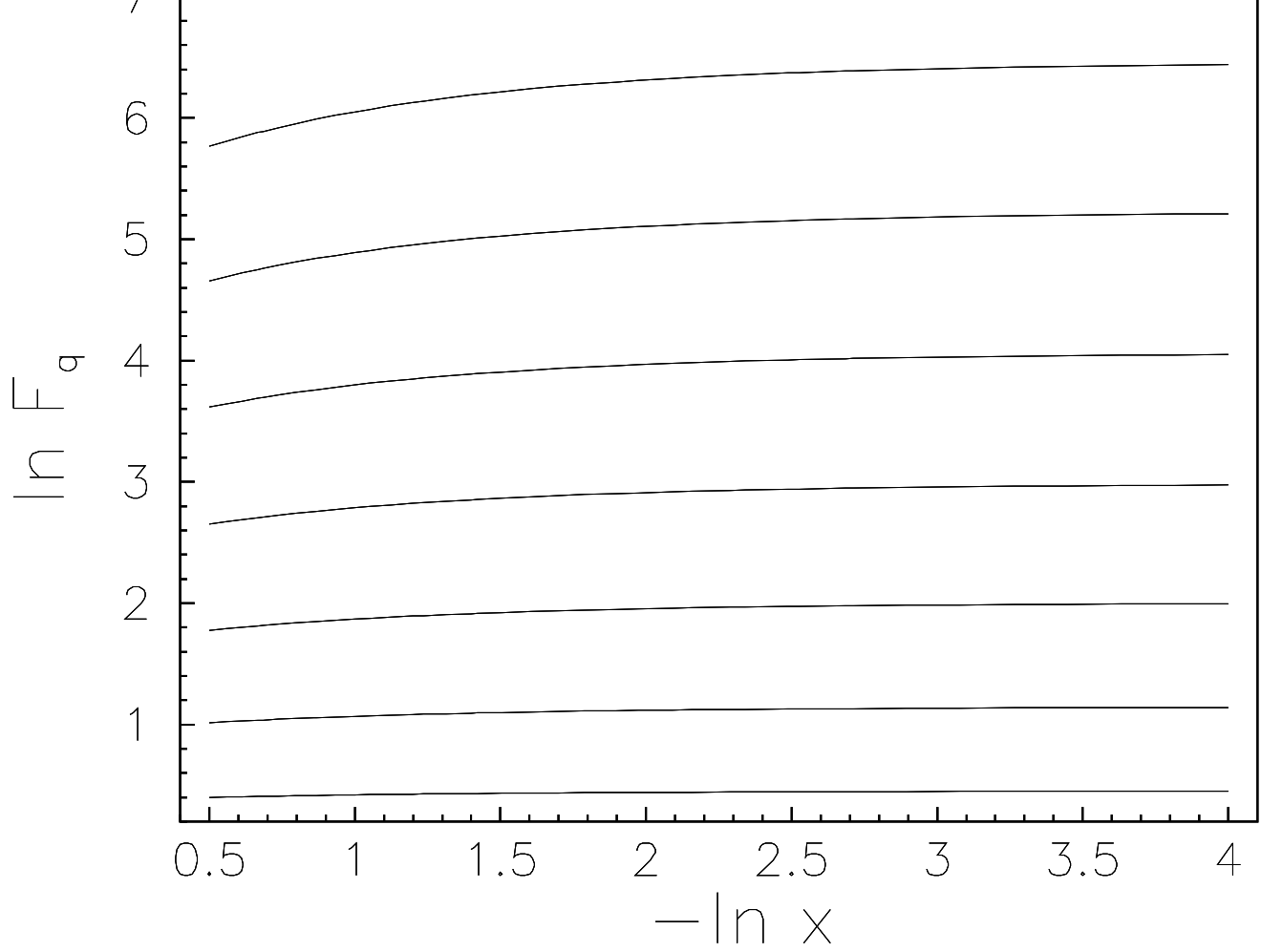
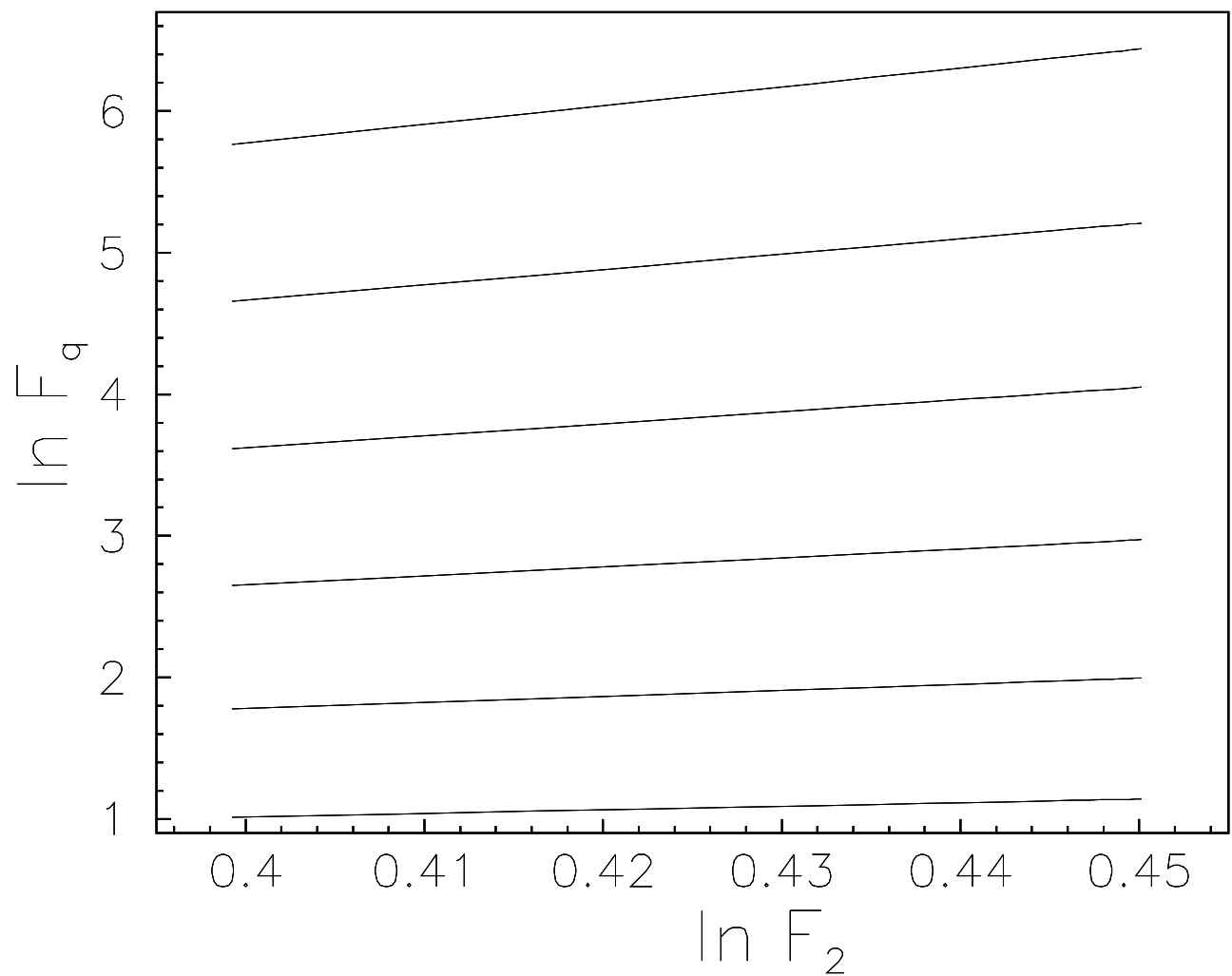
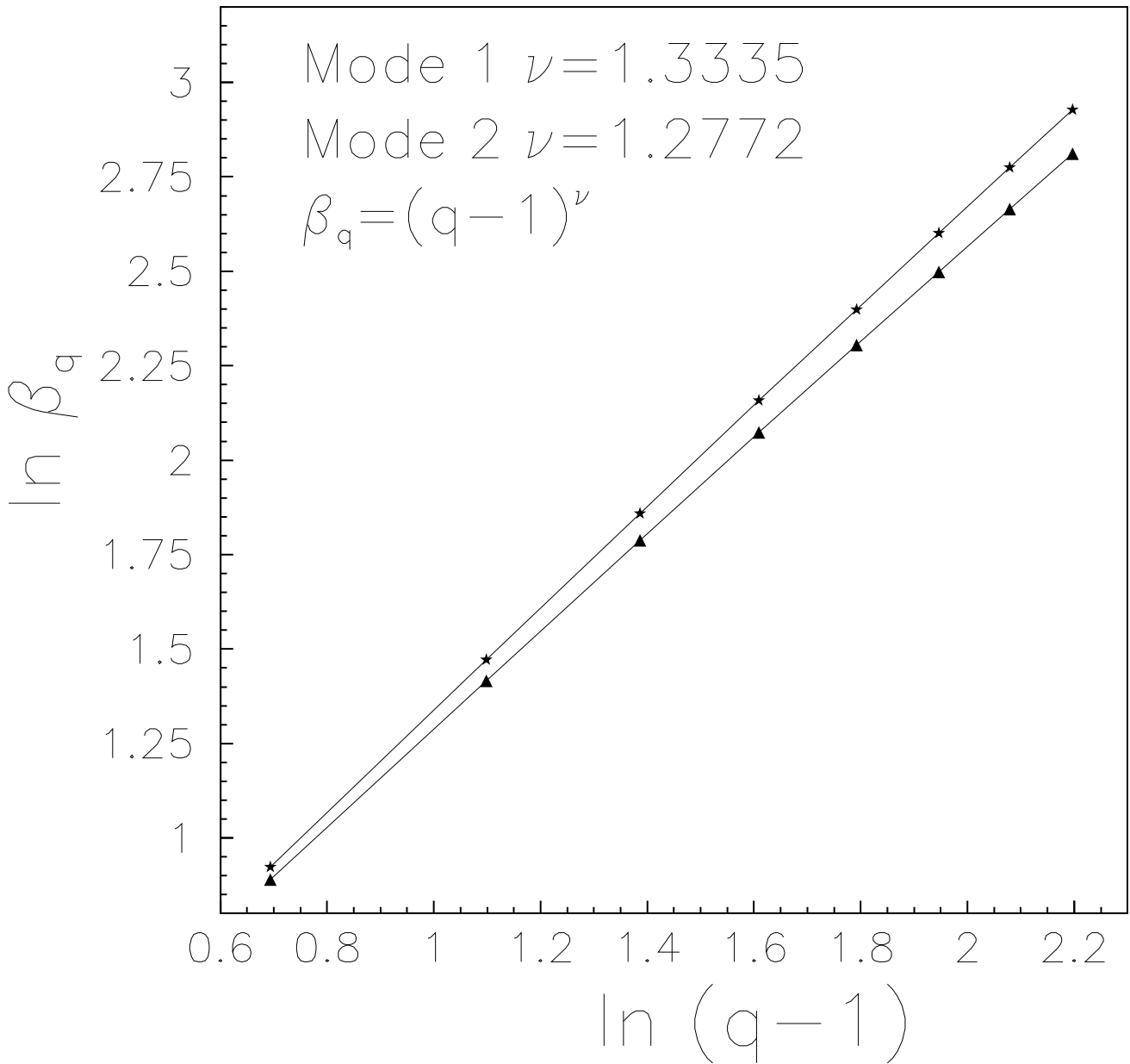
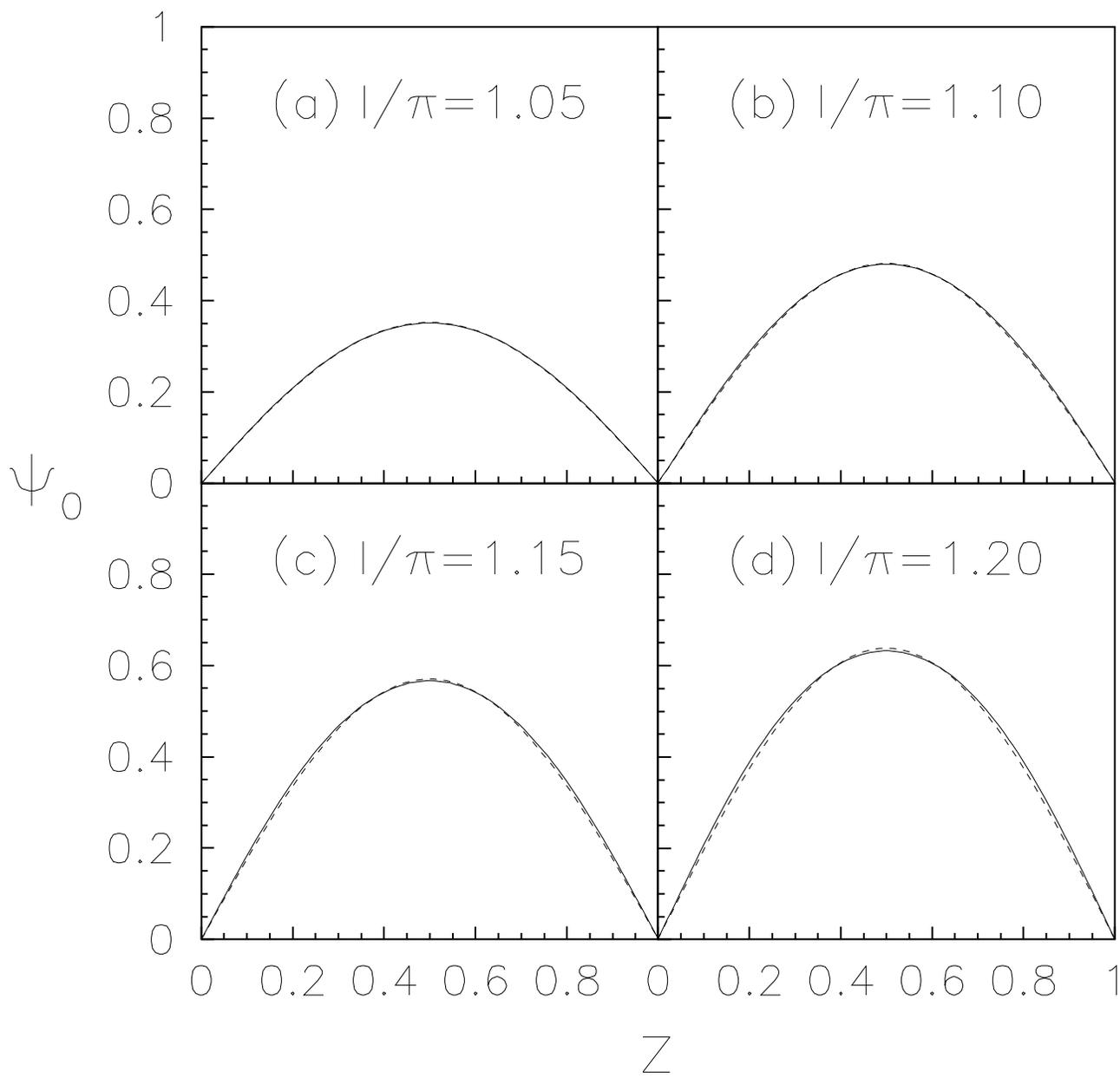
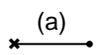
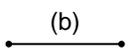
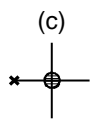
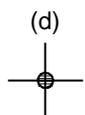
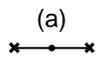
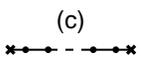
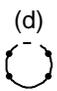
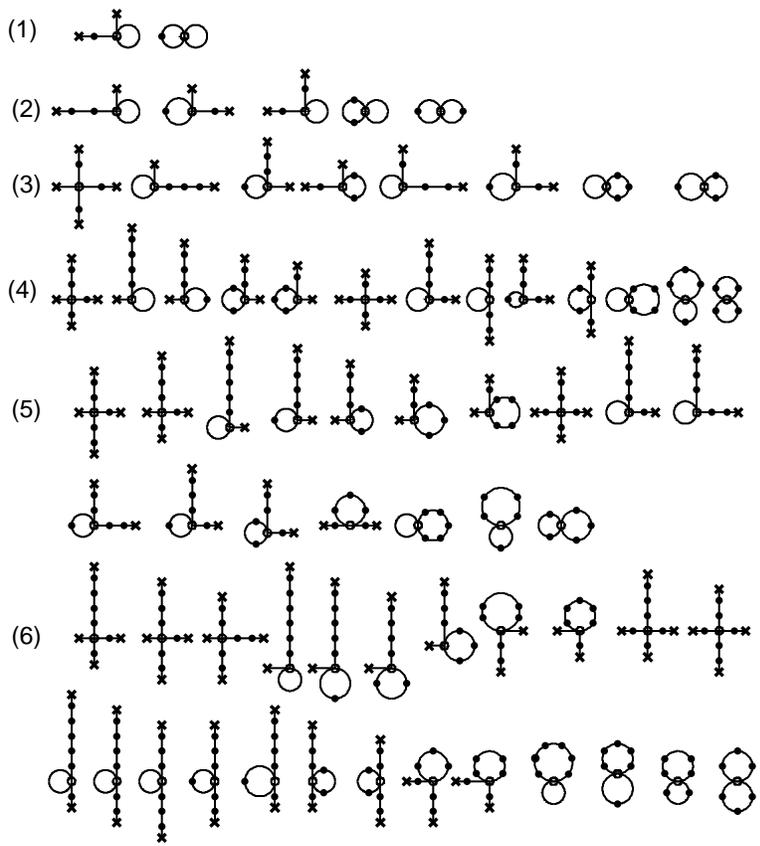
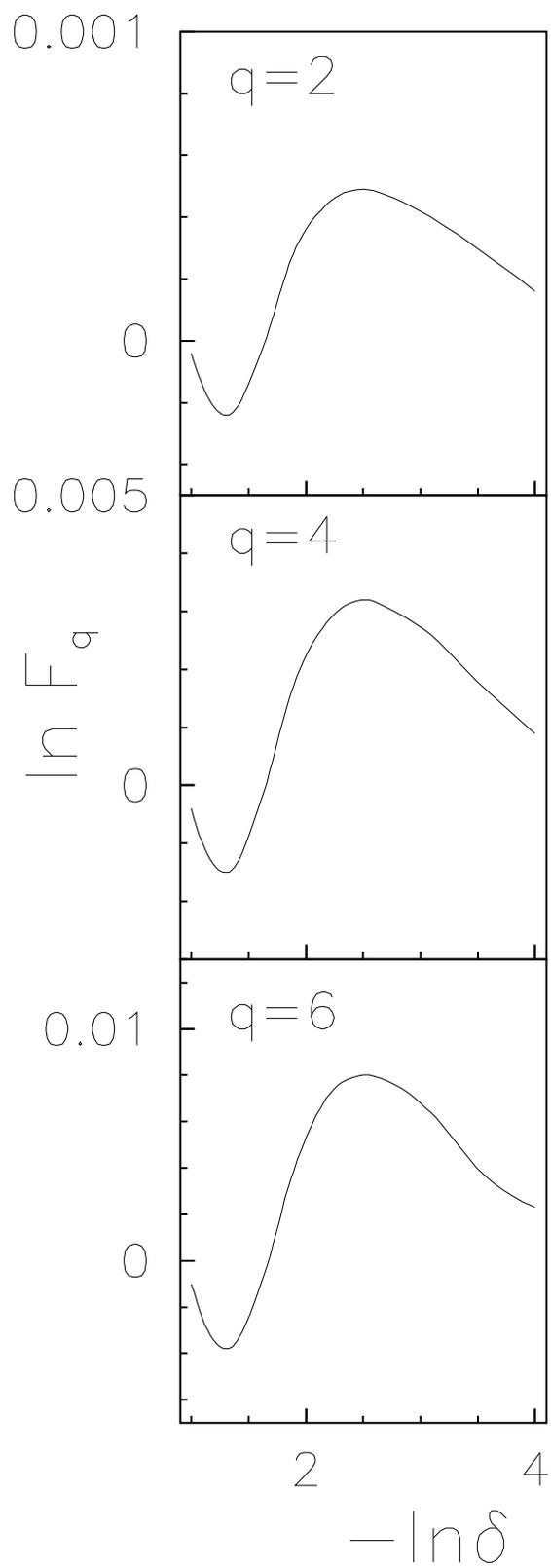
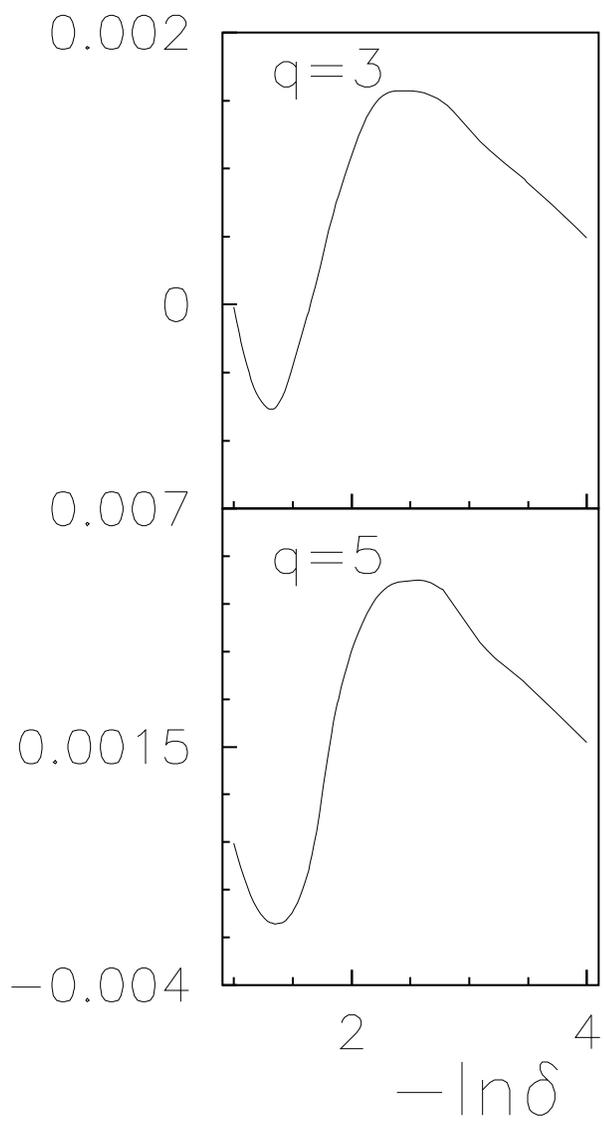
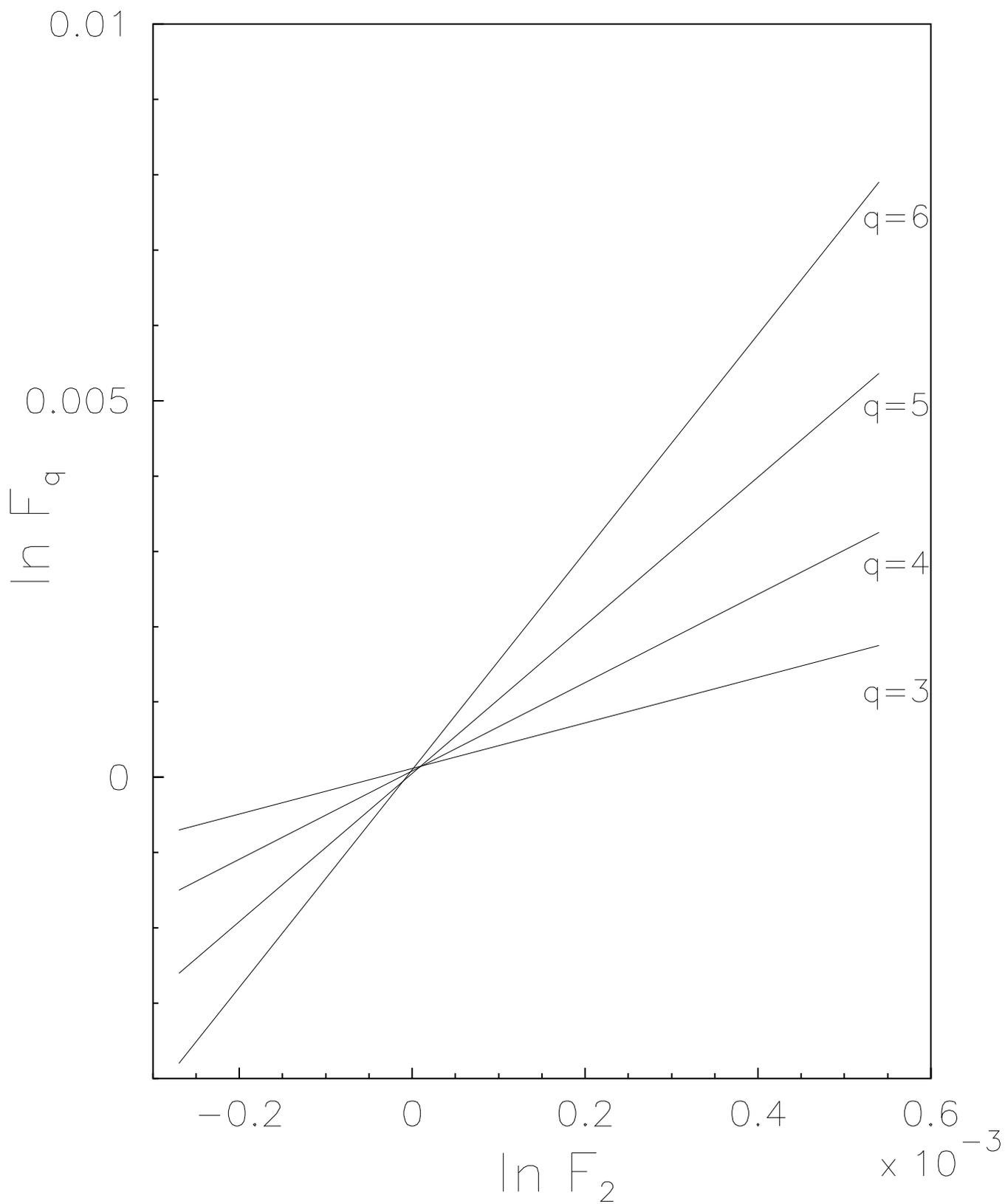
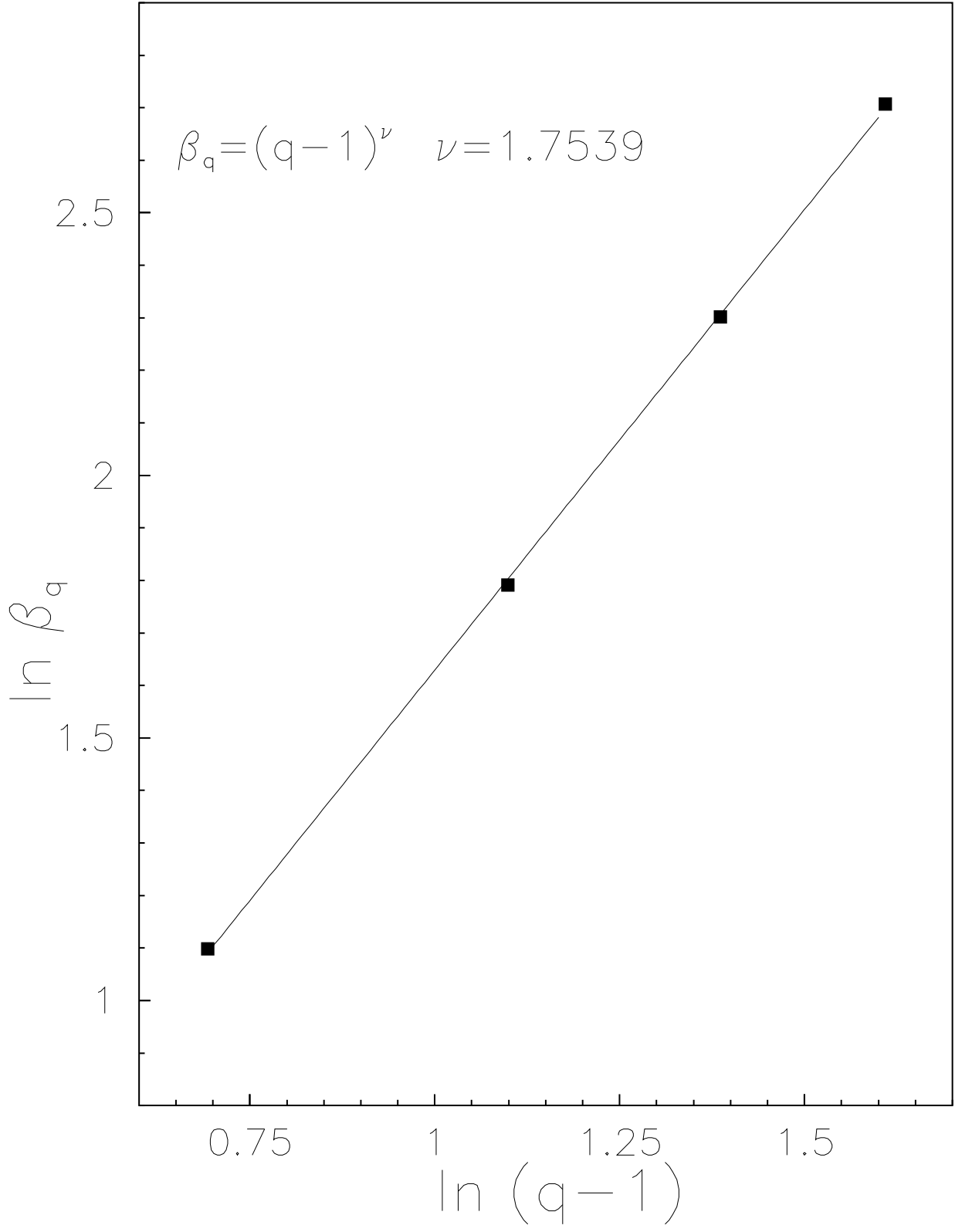
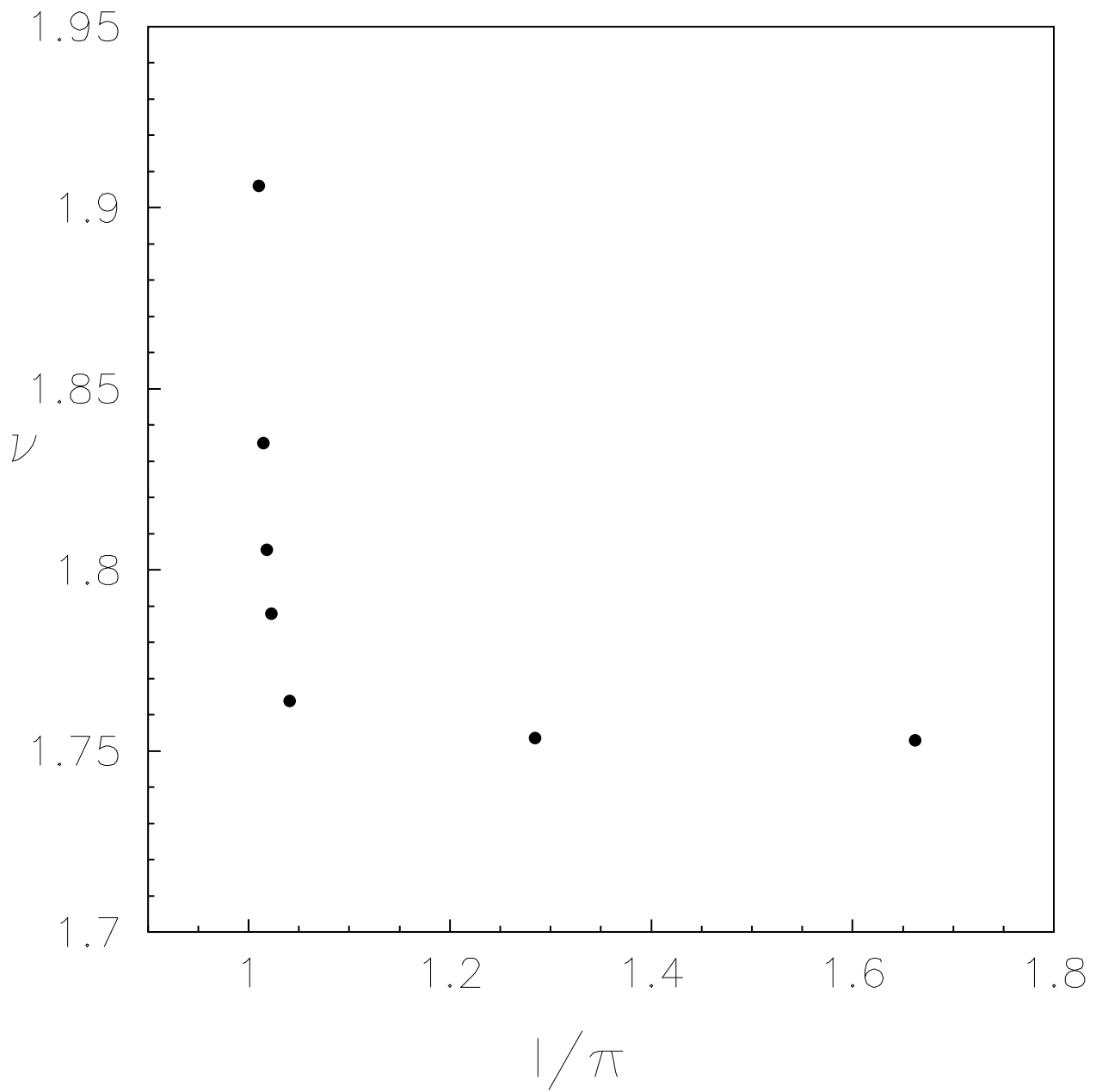
\end{document}